\documentclass[lettersize,journal]{IEEEtran}
\usepackage{amsmath,amsfonts}
\usepackage{algorithmic}
\usepackage{algorithm}
\usepackage{array}
\usepackage[caption=false,font=normalsize,labelfont=sf,textfont=sf]{subfig}
\usepackage{textcomp}
\usepackage{stfloats}
\usepackage{url}
\usepackage{verbatim}
\usepackage{graphicx}
\usepackage{makecell}
\usepackage{cite}
\usepackage[numbers,sort&compress]{natbib}
\usepackage{fancyhdr}
\begin{document}

\title{Conflict Mitigation Framework\\ and Conflict Detection in O-RAN Near-RT RIC}

\author{Cezary~Adamczyk, Adrian~Kliks
\thanks{C. Adamczyk and A. Kliks were with Poznan University of Technology, Poznań, Poland, e-mails: cezary.adamczyk@doctorate.put.poznan.pl, adrian.kliks@put.poznan.pl. The work has been realized within the project no. 0312/SBAD/8163 funded by the Poznan University of Technology.}
}

\fancyfoot[C]{\small Copyright © 2023 IEEE. Personal use is permitted, but republication/redistribution requires IEEE permission.}
\fancyhead[L]{\small This is the author’s version of an article that has been published in the IEEE Communications Magazine. Changes were made to this version by the publisher prior to publication, the early access version of the record is available at: 10.1109/MCOM.018.2200752. }

\IEEEpubid{Copyright © 2023 IEEE. Personal use is permitted, but republication/redistribution requires IEEE permission.}

\maketitle

\begin{abstract}
The steady evolution of the Open RAN concept sheds light on xApps and their potential use cases in O\mbox{-}RAN-compliant deployments. There are several areas where xApps can be used that are being widely investigated, but the issue of mitigating conflicts between xApp decisions requires further in-depth investigation. This article defines a conflict mitigation framework (CMF) built into the existing O\mbox{-}RAN architecture; it enables the Conflict Mitigation component in O\mbox{-}RAN's Near-Real-Time RAN Intelligent Controller (Near\mbox{-}RT RIC) to detect and resolve all conflict types defined in the O\mbox{-}RAN Alliance's technical specifications. Methods for detecting each type of conflict are defined, including message flows between Near\mbox{-}RT RIC components. The suitability of the proposed CMF is proven with a simulation of an O-RAN network. Results of the simulation show that enabling the CMF allows balancing the network control capabilities of conflicting xApps to significantly improve network performance, with a small negative impact on its reliability. It is concluded that defining a unified CMF in Near\mbox{-}RT RIC is the first step towards providing a standardized method of conflict detection and resolution in O\mbox{-}RAN environments.
\end{abstract}


\section{Introduction}
\IEEEPARstart{T}{he} concept of open radio access networks (O-RAN) has been widely investigated in recent years by large organizations comprising mobile network operators (MNOs), telecommunications companies, and research facilities. One of such organizations is the O\mbox{-}RAN Alliance, which formulates technical specifications to guide the industry towards O-RAN. Its efforts enable researchers and companies to build upon the laid foundations, with each new series of technical specifications providing details of the ideas that drive the whole O\mbox{-}RAN concept~\cite{oran_wg1_arch}.

The openness of O\mbox{-}RAN enables novel approaches to operating RAN and fulfilling new deployment scenarios. First of all, the interfaces between logical components of O\mbox{-}RAN are standardized and open. This enables interchangeability of O\mbox{-}RAN components manufactured by various vendors without limitation of network functionality. This is a significant change compared to traditional ``proprietary'' RANs, where RAN hardware and software are usually provided by a single vendor. Freeing MNOs from this dependency largely expands the possibilities of network evolution.

The next unique aspect of O\mbox{-}RAN is the introduction of new components that can influence the RAN operation. These include Non-Real-Time (Non\mbox{-}RT) and Near-Real-Time (Near\mbox{-}RT) RAN Intelligent Controllers (RICs), the former of which is a part of the Service Management and Orchestration (SMO) framework. These logical functions react to current and forecasted future network states and adapt its parameters to optimize performance.

\IEEEpubidadjcol

Last, but not least, the accessibility of O\mbox{-}RAN is accelerated with the concept of applications in RICs. These applications run in the Non\mbox{-}RT RIC (called rApps) and in the Near\mbox{-}RT RIC (called xApps). Both rApps and xApps can be developed by independent parties and then deployed on all O\mbox{-}RAN-compliant platforms, regardless of their hardware and software vendor. The aim of rApps and xApps is to influence the network operation from the RICs. For example, an rApp can realize a use case to adjust radio resource allocation policies to reduce latency and improve performance in dynamic handover for V2X scenarios~\cite{oran_wg1_uc_report}; the Non\mbox{-}RT RIC then interfaces with the Near\mbox{-}RT RIC to provide policy updates provided by rApps. An example of an xApp is an application that manages radio resources in such a way as to maximize Quality of Service (QoS) for a group of users~\cite{oran_wg1_uc_report} using a dedicated interface to send control messages to RAN nodes.

Prioritization within all areas of the O\mbox{-}RAN space is decided upon in the Minimum Viable Plan (MVP) towards the commercialization of the concept~\cite{oran_mvp}. It comprises key end-to-end use cases that are applicable in commercial RAN, priorities of which are chosen by MNO members of O\mbox{-}RAN Alliance. In June 2021, the priority was set on use cases such as traffic steering, QoS optimization, or Massive Multiple-Input Multiple-Output (mMIMO) optimization. Since then, multiple new use cases for O\mbox{-}RAN have been defined, with energy saving being one of the most important new focuses for network operators~\cite{oran_wg1_uc_report}. All of these use cases are to be fulfilled using various xApps and rApps working concurrently in the RICs. As multiple applications can be deployed and activated in the network at the same time, their influences on the network operation may conflict with each other. Such conflicts need to be detected and mitigated in both Non-RT and Near-RT domains.

This article proposes a novel conflict mitigation framework (CMF) to efficiently detect and resolve conflicts between network control decisions made by xApps in Near\mbox{-}RT RICs. The key contributions of this work are as follows: first, we present a comprehensive framework for conflict mitigation in O-RAN, which monitors network events to detect and resolve all conflict types between xApps, and which we propose to make standardized. Secondly, based on that framework, we propose procedures to mitigate the conflicts, along with detailed templates of exchanged messages. Finally, we propose a pragmatic way of resolving conflicts between same-importance (i.e., of the same level in the hierarchy) RAN control decisions based on prioritization and provide simulation results that show how this approach influences the network. Lastly, we outline the expected next steps in research of conflict mitigation in O\mbox{-}RAN.

\section{Conflicts between RAN control decisions}
Although O\mbox{-}RAN's flexibility and novel intelligent RAN components allow for robust RAN optimization, these features also create unique challenges in controlling the network. Control conflicts are possible as multiple agents can decide on RAN operation parameters. These conflicts can potentially nullify or lessen the decisions' influence on the network performance, so they should be avoided where possible with proper network setup~\cite{understanding_oran}. However, the complex dependencies between RAN control activities and potential expansion of network control measures make complete mitigation of conflicts with static configuration not viable. Thus, any RAN control conflicts that appear during O-RAN network operation need to be detected and resolved. For that purpose, Conflict Mitigation (CM) components are envisioned in the Near\mbox{-}RT RIC and in the Non\mbox{-}RT RIC~\cite{oran_wg3_nRT-RIC-arch, oran_wg2_NRT-RIC-arch}. Although the O-RAN architecture sections out the CM logic only in the RICs and it is the main focus of this article, some CM capabilities could be realized on the SMO level (and outside of the Non\mbox{-}RT RIC).

\begin{figure}[!t]
\centering
\includegraphics[width=0.46\textwidth,angle=0]{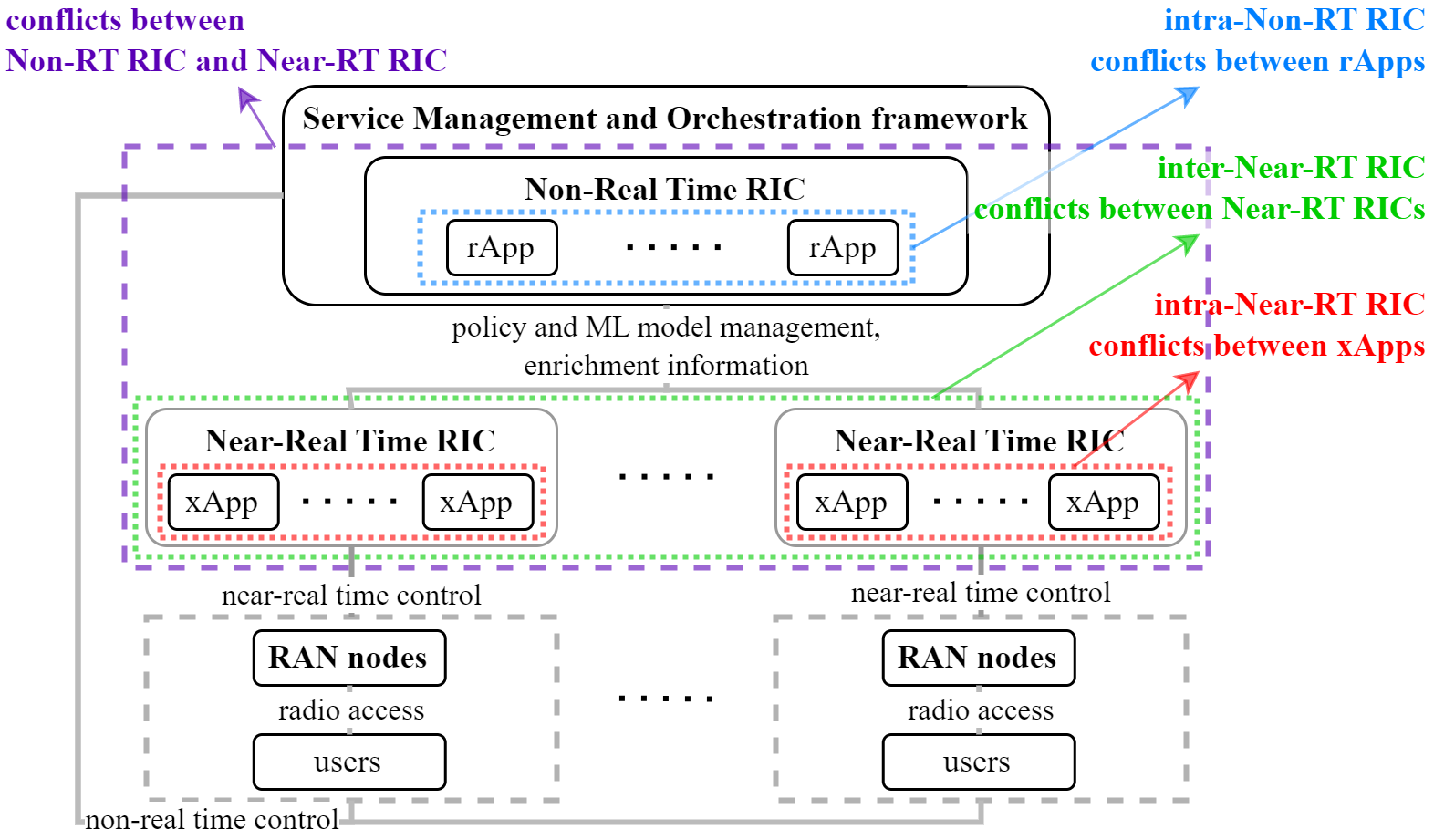}
\caption{Areas of potential control conflict in O-RAN architecture}
\label{fig_conflicts}
\end{figure}

Control decision conflicts can happen at many levels in the O-RAN architecture. Within RICs, these conflicts can happen between xApps in a Near\mbox{-}RT RIC and between rApps in the Non\mbox{-}RT RIC when control activities of applications contradict each other. Other than that, Near\mbox{-}RT RICs may also conflict each other, specifically if they control RAN nodes located in close proximity. These conflicts between equivalent components can be referred to as \textit{horizontal conflicts}. Control conflicts can also happen between components on different levels of the architecture, i.e., between the Non\mbox{-}RT RIC and Near\mbox{-}RT RICs. An example of such a type of conflict is when an xApp either acts against a policy, or contradicts a control decision provided by the Non\mbox{-}RT RIC. These conflicts can be named \textit{vertical conflicts}. All the considered areas of potential conflict within the O-RAN architecture are shown in Fig. \ref{fig_conflicts}.

The decision conflicts within O\mbox{-}RAN are considered in this article in the context of horizontal conflicts between xApps in a Near\mbox{-}RT RIC (\textit{intra-Near\mbox{-}RT RIC}) and between Near\mbox{-}RT RICs (\textit{inter-Near\mbox{-}RT RIC}). Similar considerations can be applicable for horizontal conflicts between rApps (\textit{intra-Non\mbox{-}RT RIC}), as they provide policy and non-real-time control decisions similarly to how xApps provide near-real-time control decisions. As for the vertical conflicts between the Non\mbox{-}RT RIC and Near\mbox{-}RT RICs, the Near\mbox{-}RT RIC's policy-conflicting control decisions may either be rejected or allowed by the Non\mbox{-}RT RIC. The specific conflict resolution action should be decided by the Non\mbox{-}RT RIC based on the affected use case and how the policies are enforced. The specific logic of how the Non\mbox{-}RT RIC should resolve conflicts and enforce its policies is not considered in this article.

\subsection{Types of conflicts between xApp decisions}
O\mbox{-}RAN technical specifications distinguish three types of conflicts that can happen between xApps~\cite{oran_wg3_nRT-RIC-arch}, namely direct, indirect, and implicit. The first of these are direct conflicts, which concern contradicting decisions that happen one after the other and affect the same set of configuration parameters. An example of such type of conflict may be when xApp \#1 decides to assign a user to a cell and right after that xApp \#2 decides to assign the same user to another cell. In this particular scenario, the result of the conflicting actions will be that only the decision of xApp \#2 will have an effect. Undetected direct conflicts may lead xApp \#1 to draw wrong conclusions about the effect of its decision.

Another type of conflict in the Near\mbox{-}RT RIC is the indirect conflict. It refers to a situation where decisions contradict each other when modifying parameters that influence the same areas of the RAN operation. For example, indirect conflicts can happen between xApps modifying parameters that influence handover boundaries, such as when xApp \#1 controls an antenna's electrical tilt and xApp \#2 changes the Cell Individual Offset (CIO). This may lead to situations where the effective handover boundary fluctuates heavily due to xApps reacting to each other's contradicting decisions.

The implicit conflict considers situations in which conflicting influences of network control decisions are difficult to observe and determine. Implicit conflicts are expected when many xApps simultaneously optimize the RAN operation with separate goals, controlling different network parameters. They can appear in a situation where xApp \#1 aims to maximize QoS for a group of users while xApp \#2 attempts to minimize the number of handovers between neighboring cells; decisions made by these two xApps may interfere in a non-obvious manner, disrupting each other's effect on the network.

Conflicts may also happen between Near\mbox{-}RT RICs that control RAN nodes in close proximity (i.e., the inter-Near\mbox{-}RT RIC conflicts). In such cases, the conflicting Near\mbox{-}RT RICs provide contradicting RAN control decisions, e.g., triggering a ``ping-pong'' handover of a user between cells managed by both Near\mbox{-}RT RICs. Such network behavior leads to inefficient radio resource utilization.

\subsection{Dealing with conflicts and related works}
To achieve predictable performance and behavior of O\mbox{-}RAN, it is necessary to minimize any negative impact of conflicts between xApp decisions. As prevention is preferred, the design of Near\mbox{-}RT RIC implementations should aim to avoid any conflicts between xApps. Unfortunately, some conflicts will inevitably happen with various third parties providing xApps. Therefore, the Near\mbox{-}RT RIC must be able to resolve any conflicts between xApp decisions. To do so, the conflict firstly needs to be detected. Currently, no methods of detecting or resolving conflicts are defined in O\mbox{-}RAN's technical specifications as they are considered for future study~\cite{oran_wg3_nRT-RIC-arch}.

The topic of conflict mitigation in O\mbox{-}RAN has not yet been significantly covered in research papers. One of the relevant papers on a related topic has been published by Zhang et al.~\cite{team_learning}, who proposed a team learning algorithm based on Deep Q-learning that requires two xApps working in cooperation in an O\mbox{-}RAN environment. While the referenced work does not describe a universal solution to the issue of conflicts in O-RAN, it proposes a viable way to mitigate the conflicts using a cooperative machine learning scheme. Results of the team learning algorithm showed that when xApps take into consideration each other's decisions, the learning process is much more efficient in comparison to xApps working independently, leading to overall higher system throughput and lower packet drop rate. Although promising, this approach is hard to scale to a larger number of xApps because the decision logic of all participating xApps needs to adapt to the number of cooperating xApps and be able to properly interpret their decisions. Therefore, the team learning approach can be utilized only if xApps are designed with team learning in mind.

Significant work has been done in the area of conflict mitigation in the context of Self-Organized Networks (SON). The concept of SON has been introduced as part of Long Term Evolution (LTE) networks to reduce operating costs related to multi-vendor RAN. SON allows for the deployment of SON functions, which, as xApps and rApps in O\mbox{-}RAN, are able to modify network behavior, albeit only locally. Two SON functions widely considered in research are Mobility Robustness Optimization (MRO) and Mobility Load Balancing (MLB). Relevant work on conflict mitigation in the context of SON includes a scheme to resolve conflicts between MRO and MLB, which limits the range of changes to CIO values done by MLB~\cite{conflict-avoidance_MRO-MLB_2010}. This approach has been improved in later research to find optimal values of CIO instead of limiting the range~\cite{conflict-avoidance_MRO-MLB_2018}. To address the problem of decision conflicts in a more general way, a soft classification of conflicts in SON has been proposed~\cite{framework-classification-SON_2013}. A general framework for self-coordination in SON has been defined to universally deal with conflicts, regardless of conflicting SON function types~\cite{coord_framework_self-organisation_2011}; it describes the concept of SON Coordinator that adapts network control to current conditions, mitigating any conflicts and providing safeguards from network performance degradation. Another generalized solution in SON utilizes machine learning to learn from past experience and predict network performance to better solve the conflicts~\cite{conflict-resolution_self-coord_2018}.
Although RAN control problems in SON and O-RAN share many similarities (i.e., SON functions and xApps/rApps as conflicting agents, common use cases), the solutions for SON cannot be directly applied to O-RAN. This is mainly due to a significant difference in how RAN control is realized in SON and O-RAN. SON functions perform RAN control locally (i.e., each RAN node optimizes its parameters based on what it can observe), while RAN control in O-RAN is executed with RICs, which can observe the entire managed network and optimize it centrally. Nevertheless, the relevant work on SON can provide insight into how the problem of conflict mitigation can be approached and what measures may prove effective in solving the control conflict problems.

Finally, it is worth mentioning multi-agent reinforcement learning (MARL) methods that consider multi-agent systems (MAS), where autonomous agents with various goals perform actions in a shared environment \cite{Nowe2012}. In an MAS, agents' actions influence each other and need to be coordinated to reach an optimal state of the environment. This case can be considered analogous to the O-RAN scenario, where xApps and rApps provide RAN control decisions to fulfill various use cases. According to the O-RAN architecture descriptions provided by O-RAN Alliance, the default RAN control setting in O-RAN is centralized and features the RICs as coordinators of RAN control activities. Hence, cooperative and coordinated MARL approaches could be applied. Although this is a viable solution, it requires xApps and rApps to be implemented with MARL in mind. We consider this a potential evolution path for the conflict mitigation methods in O-RAN. However, this article focuses on solutions that can be applied to all applications, regardless of their implementation.

\section{Conflict Mitigation Framework}
\begin{figure}[!t]
\centering
\includegraphics[width=0.46\textwidth,angle=0]{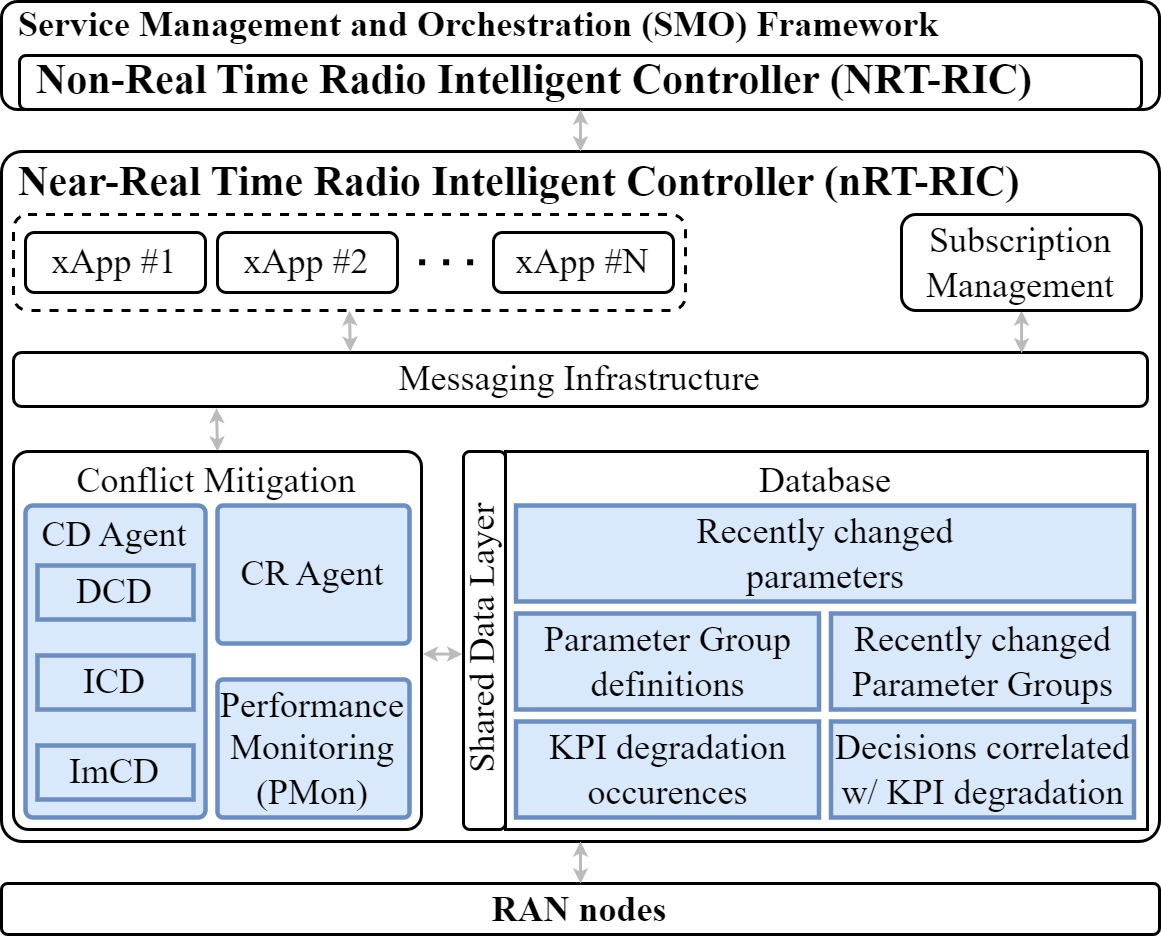}
\caption{CMF components embedded within the Near\mbox{-}RT RIC architecture}
\label{fig_cdf}
\end{figure}

In this article, we propose a framework for conflict mitigation (referred to in the paper as CMF) in O\mbox{-}RAN-compliant environments. The proposed framework is envisioned as part of the CM component of Near\mbox{-}RT RIC. CMF aims to provide a robust, standardized method for conflict detection in O\mbox{-}RAN, covering the detection of all three types of conflict defined in O\mbox{-}RAN specifications. The proposed framework embedded within the architecture of Near\mbox{-}RT RIC is shown in Fig.~\ref{fig_cdf}.

\subsection{Structure of CMF and infrastructure prerequisites}
CMF envisions deploying the \textit{Conflict Detection} (CD) \textit{Agent} and the \textit{Conflict Resolution} (CR) \textit{Agent} components in the Near\mbox{-}RT RIC's CM entity as agents that detect conflicts between xApps and resolve them. The main research challenge for the CD Agent is to enable reliable detection of all types of conflicts, and for the CR Agent to enable the resolution of all conflicts to reach an optimal network state. There are three logical parts distinguished within the CD Agent that are dedicated to detection methods for each type of conflict between xApp decisions. These components are described in the following subsections. As for the CR Agent, it is assumed that it can resolve any conflicts detected by the CD Agent.

The operation of CMF relies on the configuration of the O-RAN infrastructure. Data stored by CMF is kept in the Near\mbox{-}RT RIC's Database, which is accessible via the Shared Data Layer (SDL). The Message Infrastructure in the Near\mbox{-}RT RIC needs to be configured to redirect all control messages from xApps into the CM component, so that all RAN control messages can be evaluated by CMF. Depending on the results of CMF's evaluation of the specific control messages, these messages can either be allowed to modify the network configuration (with or without modifications, depending on the logic implemented in the CR Agent) or be entirely blocked from influencing the network. It should be acknowledged that adding an additional step in the RAN control message processing pipeline increases the delay between an xApp providing a RAN control message and the message influencing the network. As Near\mbox{-}RT RIC aims to control the network operation in near-real-time scale, any processing done in the CM component, i.e., in the CD Agent and the CR Agent, needs to be relatively quick. Another dependency on the Near\mbox{-}RT RIC architecture is that the CD Agent needs to be subscribed to performance management (PM) data provided by RAN nodes managed by the Near\mbox{-}RT RIC. This is required for the detection of implicit conflicts, which is described in details in a subsequent section.

To mitigate the inter-Near\mbox{-}RT RIC conflicts, CMF envisions the exchange of information between Near\mbox{-}RT RICs about RAN control messages that are currently in effect, via the Non\mbox{-}RT RIC. This information shall be a part of the enrichment information interface between the Non\mbox{-}RT RIC and Near\mbox{-}RT RICs. Once the Near\mbox{-}RT RICs are aware of all control decisions for users, cells, and bearers managed by other Near\mbox{-}RT RICs, these decisions can be considered as part of the conflict mitigation procedures for shared control targets.

\subsection{Direct Conflict Detection}
Ideally, direct conflicts should be detected as part of pre-deployment xApp assessment, which is done by the MNO deploying the xApps. Nevertheless, a method for detection of direct conflicts is required as a fail-safe mechanism, e.g., in case of human error leading to deployment of directly conflicting xApps.

Direct conflicts can be detected by the \textit{Direct Conflict Detection} (DCD) component within the CD Agent. DCD can detect the conflicts pre-action (i.e., before a control decision is effective), as it detects changes of recently changed parameters related to a user, a cell, or a bearer. All RAN control messages are tracked by the CD Agent in the Database (``Recently changed parameters'' in Fig.~\ref{fig_cdf}) along with a timestamp, the source xApp, the control target, the parameters modified by the message, and the control time span of the message (i.e., the control duration expected by the xApp).

The conflict detection logic of DCD is as follows: each new control message in the Near\mbox{-}RT RIC is saved into the Database, then it is compared to all currently effective xApp control decisions; if any decisions share the control target and at least one of the modified parameters, data about the conflicting decisions is provided to the CR Agent. The DCD procedure is illustrated in Fig.~\ref{fig_dcd+icd_logic}, which shows the message exchange flow between the CMF components.

\subsection{Indirect Conflict Detection}
Indirect conflicts cannot be directly observed by the \textit{Indirect Conflict Detection} (ICD) component within the CD Agent, but it can anticipate them pre-action by having the knowledge of groups of parameters that influence the same area of RAN operation. These groups of parameters can be configured manually by the MNO, predefined in the standards, or learned dynamically during network operation using \textit{Performance Monitoring} (described in the next section). The groups are stored in the Database as \textit{Parameter Groups} (PGs) (``Parameter Group definitions'' in Fig.~\ref{fig_cdf}). Additionally, the Database stores information about RAN control messages that modify any parameters from these groups (``Recently changed Parameter Groups'' in Fig.~\ref{fig_cdf}), with the same scope of data as ``Recently changed parameters'' related to DCD.

\begin{figure}[!t]
\centering
\includegraphics[width=0.46\textwidth,angle=0]{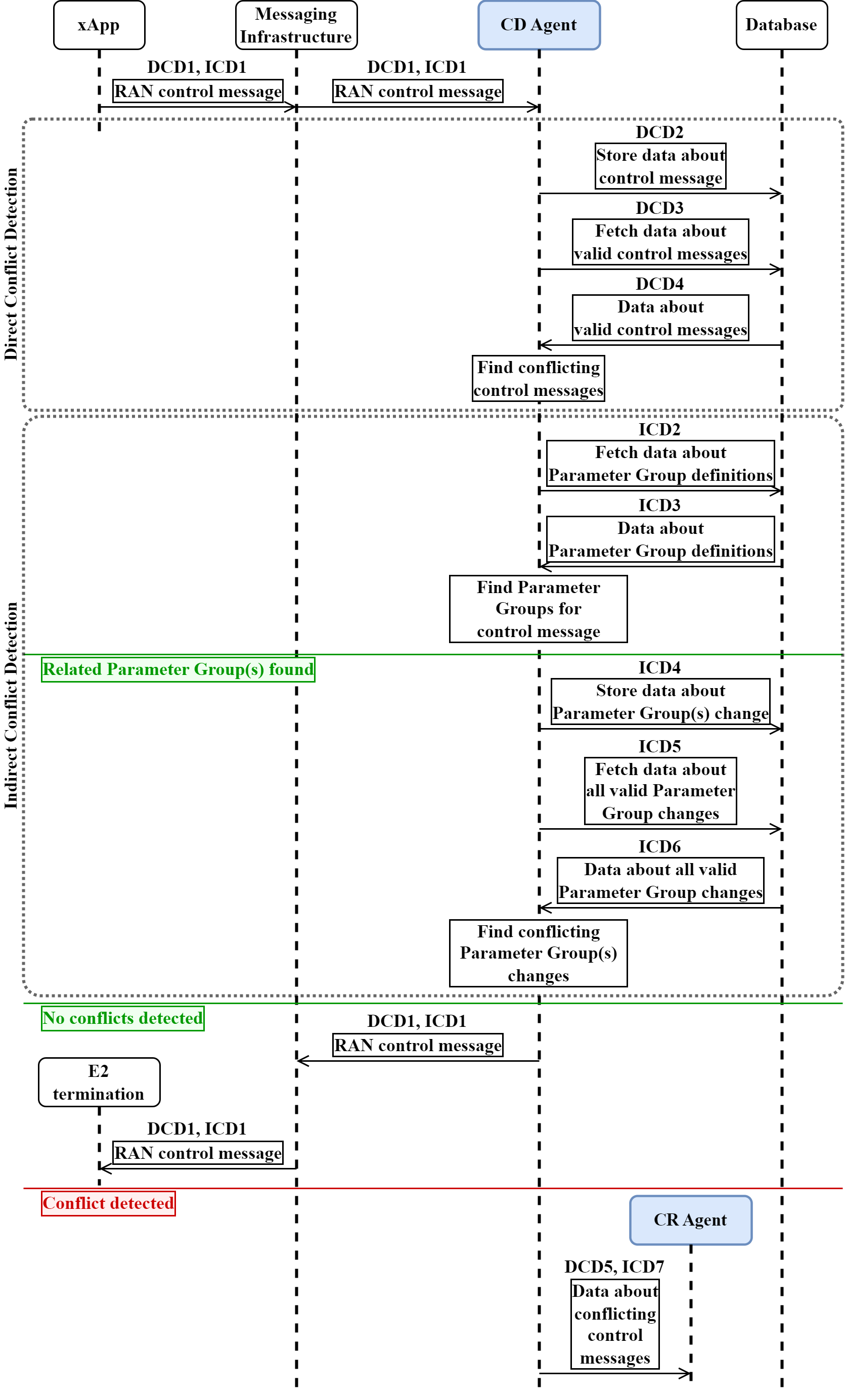}
\caption[]{Direct and Indirect Conflict Detection message flow{\footnotemark[1]}}
\label{fig_dcd+icd_logic}
\end{figure}

Detection of indirect conflicts is a slight modification of DCD: it also analyzes each new control message but first maps the target parameter onto predefined PGs. After the message is noted against at least one of the known PGs, ICD checks the entries in ``Recently changed Parameter Groups'' to find any conflicting decisions. If a match is found, information about the conflicting xApp decisions is provided to the CR Agent. The conflict detection logic and related message flow for ICD are shown in Fig.~\ref{fig_dcd+icd_logic}, alongside the DCD procedure.

\subsection{Implicit Conflict Detection}
Implicit conflicts cannot be observed directly, and the mutual influence between conflicting decisions is not easily deduced. Therefore, a similar approach, as described for DCD and ICD, is not suitable for \textit{Implicit Conflict Detection} (ImCD), which can only work reactively. An additional component within ImCD is required to analyze PM data reported by RAN nodes and detect any occurrences of network performance degradation; this role is fulfilled by the Performance Monitoring (PMon) component within CMF. Detection of network performance anomalies can be achieved with either traditional statistical-based solutions or AI/ML-based approaches \cite{ml-anomaly_2016}. The aim of the PMon component is to detect any significant RAN Key Performance Indicator (KPI) degradation. ImCD utilizes Near\mbox{-}RT RIC's Database accessed via the SDL to store data about RAN KPI degradation occurrences and KPI degradation-correlated xApp decisions.

\begin{figure}[!t]
\centering
\includegraphics[width=0.46\textwidth,angle=0]{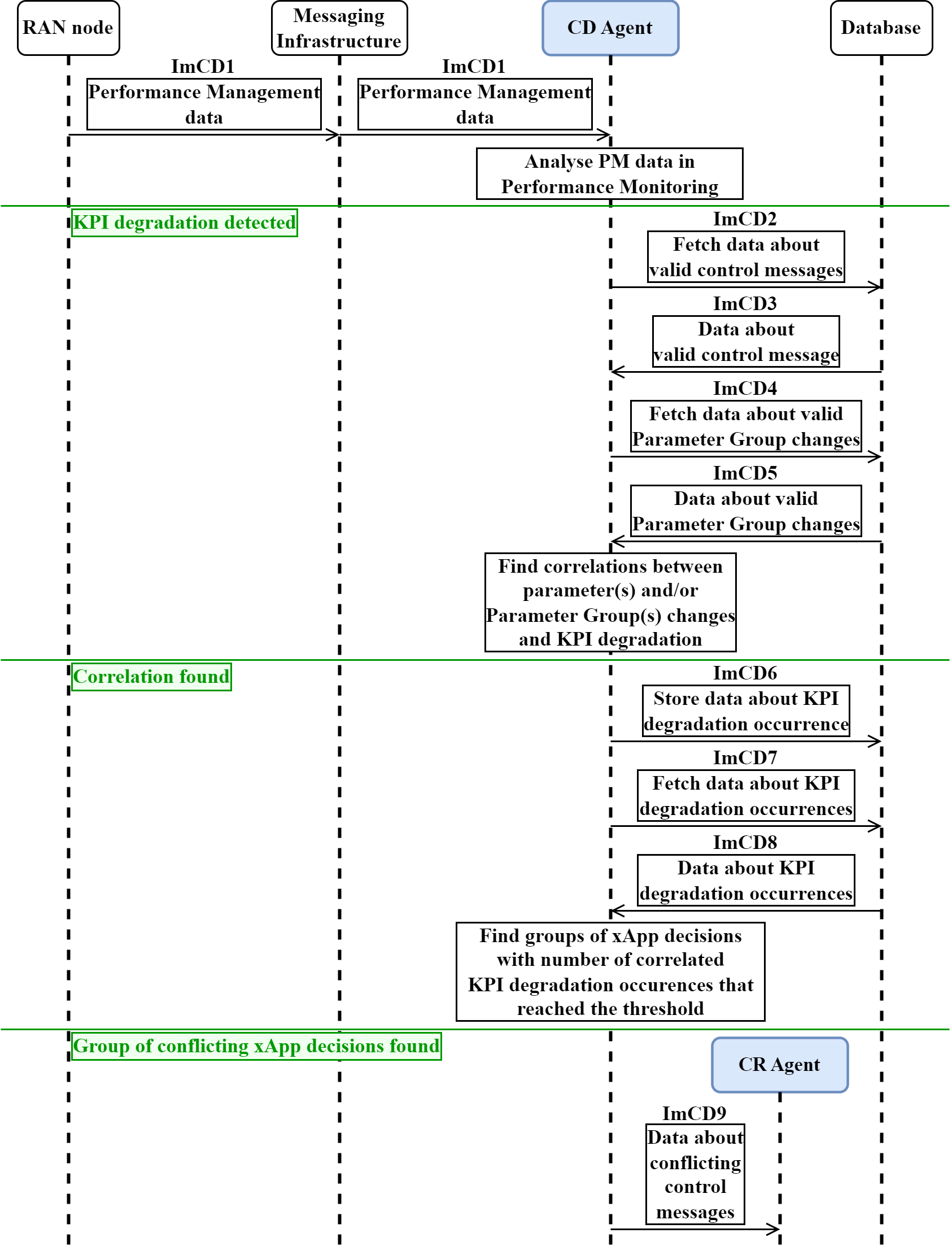}
\caption[]{Implicit Conflict Detection message flow{\footnotemark[1]}}
\label{fig_imcd_logic}
\end{figure}

The operation of ImCD is not triggered by any RAN control message but by PMon signaling a detection of RAN KPI degradation. Once the trigger is provided, ImCD analyzes which parameters and/or Parameter Groups have been recently modified by many control messages. ImCD utilizes the data captured in the Database by DCD and ICD, so it does not need to monitor any control messages by itself. If a correlation is found between any recent RAN control decisions and the observed RAN KPI degradation, ImCD notes it and increments its internal counters relevant to these control decisions. When any tracked counter breaches a predefined threshold, the CD Agent provides data about conflicting xApps' control decisions to the CR Agent. At this point in the procedure, the CD Agent may remove data about RAN KPI degradation occurrences related to the reported conflict from the Database. A significant difference between DCD/ICD and ImCD is that the latter works post-action and, therefore, cannot entirely prevent conflicting RAN control messages from influencing the network. Similar to the DCD and ICD procedures described earlier, the message exchange flow and conflict detection logic for the ImCD procedure are depicted in Fig.~\ref{fig_imcd_logic}.

\addtocounter{footnote}{+1}\footnotetext{Examples of the messages exchanged by the O-RAN components during CMF operation are available online for readers' reference (free access)~\cite{repo_adamczyk2023}. The filenames of the messages in the repository align with the labels of the message flows shown in Fig.~\ref{fig_dcd+icd_logic} and Fig. \ref{fig_imcd_logic}.}

\section{Evaluation of the solution}
\subsection{Simulation scenario}
To evaluate the efficiency of CMF, we simulated a 19-base-station O-RAN network with the base stations (BSs) evenly distributed in an urban environment on a hexagonal grid, 600 meters apart. The simulation area is limited to approximately outline the area covered by the network. There are 380 users spread across the area, each with one of three user profiles (i.e., low, medium, and high bitrate), chosen randomly with various probabilities (60\% for low, 30\% for medium, and 10\% for high bitrates). Users move randomly around the considered area, either as pedestrians or in a vehicle, generating traffic in the network. Connections can be handed over between BSs as propagation conditions change.

The simulated network includes a nRT-RIC, which monitors performance parameters for all BSs. Within the Near\mbox{-}RT RIC, two xApps are installed: MRO and MLB. MRO monitors handover statistics of each BS and modifies handover hysteresis and time-to-trigger parameters to minimize the number of radio link failures (RLFs) and ping-pong handovers. MLB balances the load of the BSs in the network, choosing CIO values according to the load of the BSs.

\begin{figure}[!t]
\centering
\includegraphics[width=0.46\textwidth,angle=0]{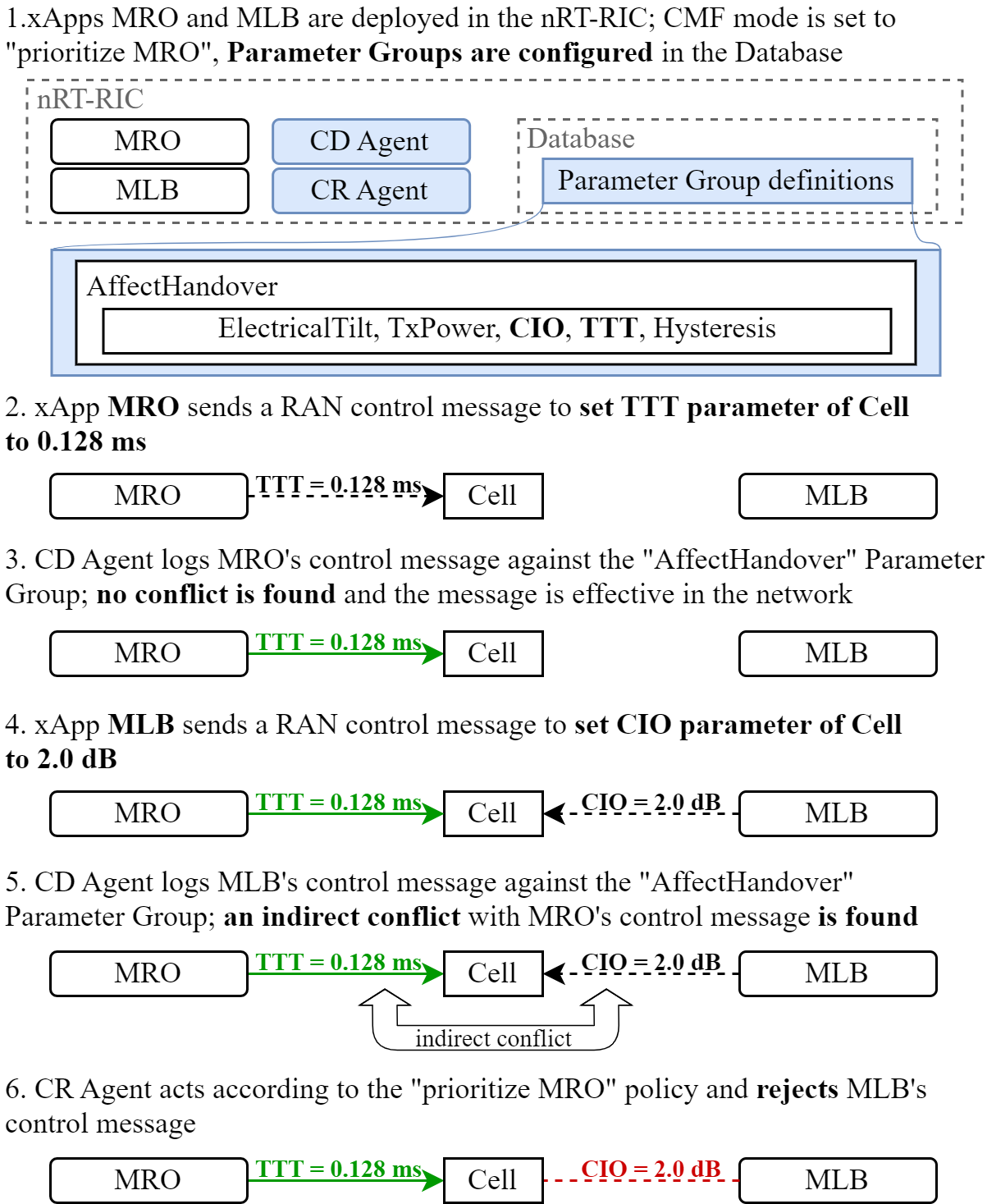}
\caption{ICD event sequence in the considered scenario}
\label{fig_icd_event_sequence}
\end{figure}

To simulate CMF's operation, CD and CR Agents are deployed within the Near\mbox{-}RT RIC. The CD Agent implements DCD and ICD, i.e., it monitors configuration modifications made by xApps and detects direct and indirect conflicts (ImCD is not applicable in the considered scenario). If no conflicts are detected, any decisions provided by xApps take effect in the order they are provided. Otherwise, conflicts are reported into the CR Agent, which decides if and when specific decisions take effect. In the considered scenario, CMF can prioritize one of the conflicting xApps. If a given xApp is prioritized, each of its decisions takes effect on the network regardless of conflicts. To showcase how CMF is able to mitigate conflicts between xApps in the considered scenario, a description of ICD operation event sequence is shown in Fig.~\ref{fig_icd_event_sequence}.
Readers interested in the details of the simulation scenario and the raw result data can access them online (free access)~\cite{repo_adamczyk2023}.

\subsection{Evaluation results}
The simulation is conducted with MRO and MLB xApps activated and one of three modes of CMF operation: disabled, prioritize MRO, prioritize MLB. The first 150 seconds out of 1,000 simulated seconds are ignored in the statistics for the sake of disregarding the initial instability of the simulated network. Tracked network performance indicators are: (i)~mean BS load, (ii)~mean user satisfaction, (iii)~number of call blockages (CBs), (IV)~number of RLFs, (V)~number of handovers, and (VI)~number of ping-pong handovers.

The performance indicators observed for the network with no conflict mitigation in place show a significant number of CBs, RLFs, and handovers, including ping-pong handovers. The excessive number of handovers causes wasteful utilization of radio resources, which, in turn, decreases the QoS for the users handled by heavily loaded BSs. In extreme cases, the BSs drop incoming traffic, therefore causing call blockages. RLFs, on the other hand, are caused by either too early or too late handovers, which result from suboptimal configuration of handover parameters.

Enabling conflict resolution with prioritization of the MRO xApp has a mostly positive effect on network performance. Although the mean load of all BSs increases slightly, the total number of handover events decreases by over 7\%. This positive impact is due to MRO being able to steer handover parameters without constraints, while changes in network configuration done by MLB are significantly limited. On the other hand, in case of conflict with MRO, the MLB xApp is unable to increase the CIO of heavily loaded BSs, which would otherwise relieve their load and free up their radio resources for incoming calls. This limitation leads to an increase in the number of CBs because users trying to connect to heavily loaded BSs may not receive proper service due to the lack of available radio resources.

The influence of CMF on network operation differs when the MLB xApp is prioritized. Mean user satisfaction increases by 0.7\%, and the number of CBs decreases by over 7\%, along with just a 0.1\% increase in mean BS load. However, all other tracked network performance indicators deteriorate to varying degrees. CMF limits the RAN control capabilities of the MRO xApp, preventing it from optimizing handover parameters in case of conflict with MLB. This leads to more occurrences of all handovers, including too late and too early handovers, which cause RLFs.

\begin{figure}[!t]
\centering
\includegraphics[width=0.46\textwidth,angle=0]{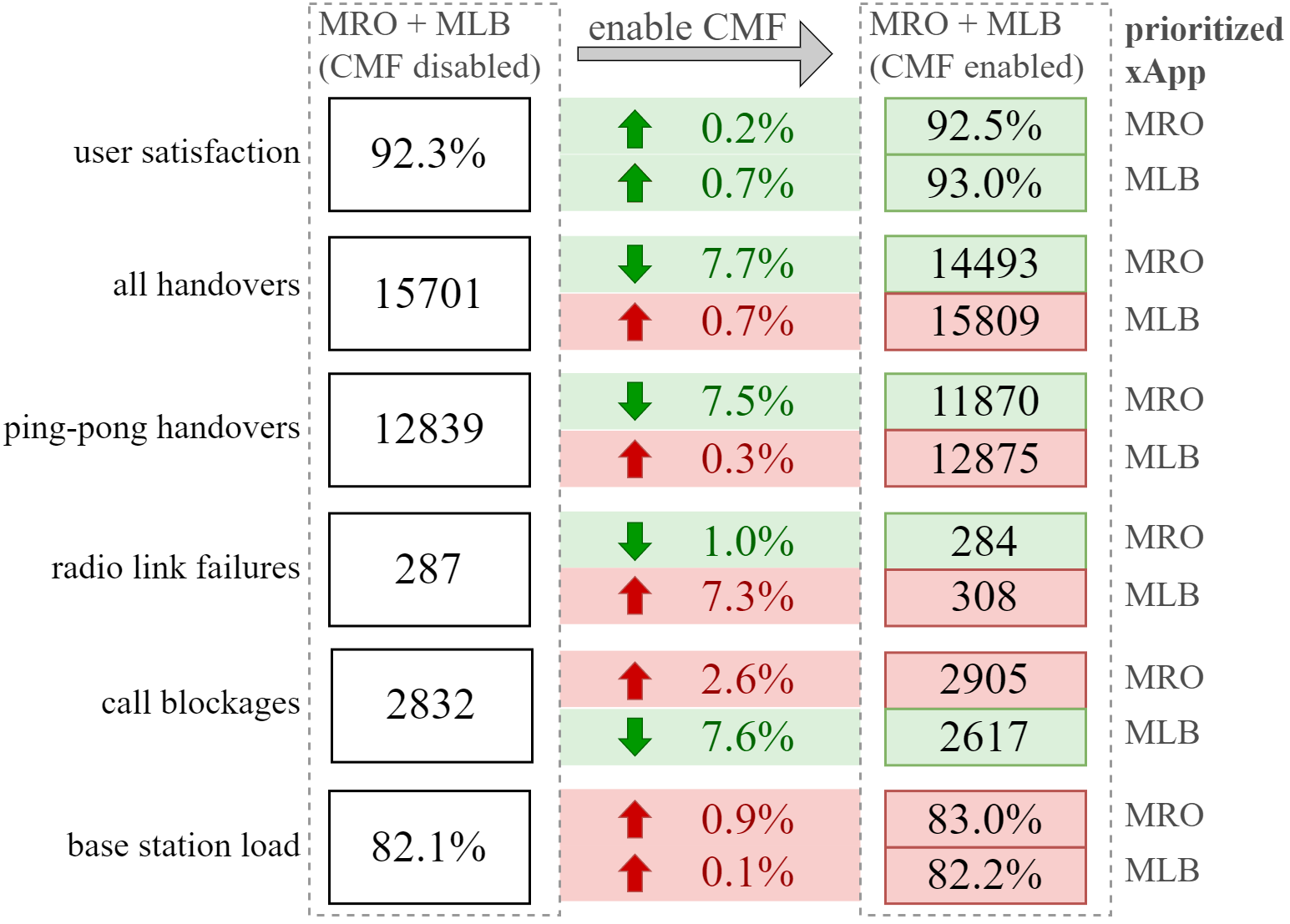}
\caption{Simulation results showing performance influence of CMF}
\label{fig:sim_results}
\end{figure}

In conclusion, the ``prioritize MRO'' mode of CMF has a more positive impact on the network compared to the ``prioritize MLB'' mode, as it causes lesser deterioration of some KPIs in exchange for its performance improvements. CMF's influence on KPIs in both modes is shown in Fig.~\ref{fig:sim_results}.

\section{Summary}
In this article, we defined a framework for conflict mitigation in O\mbox{-}RAN's Near\mbox{-}RT RIC. By design, it is capable of mitigating all three types of conflicts described in the technical specifications published by O\mbox{-}RAN Alliance: direct, indirect, and implicit conflicts. The proposed framework works by tracking all RAN control messages sent by xApps deployed within the Near\mbox{-}RT RICs and monitoring PM data reported by RAN nodes. The concept of a conflict mitigation framework is easily scalable and does not impose any restrictions on how conflict resolution is realized. We provided a complete set of example JSON messages exchanged as part of the procedures related to the detection of all types of conflicts between xApps.

The efficiency of the CMF concept was proven with a simulation of an O\mbox{-}RAN network with and without mitigation of xApp conflicts. The simulation results show that enabling CMF, even with a basic scheme of conflict resolution, can reduce the negative effects of xApp conflicts.

Defining the CMF is envisioned as the first step in providing a standardized conflict mitigation method in the Near\mbox{-}RT RIC. We expect future work in this field to include (i)~the development of the PMon component, potentially with use of AI/ML-based tools for detection of performance degradation, (ii)~the development of the CR Agent entity, compatible with CMF, and with optimal conflict resolution logic for each type of conflict in the Near\mbox{-}RT RIC, and (iii)~open source implementations of CMF for utilization in O-RAN networks.

\bibliographystyle{IEEEtran}
\bibliography{IEEEbibliography}

\vspace{11pt}

\begin{IEEEbiography}[{\includegraphics[width=1in,height=1.25in,clip,keepaspectratio]{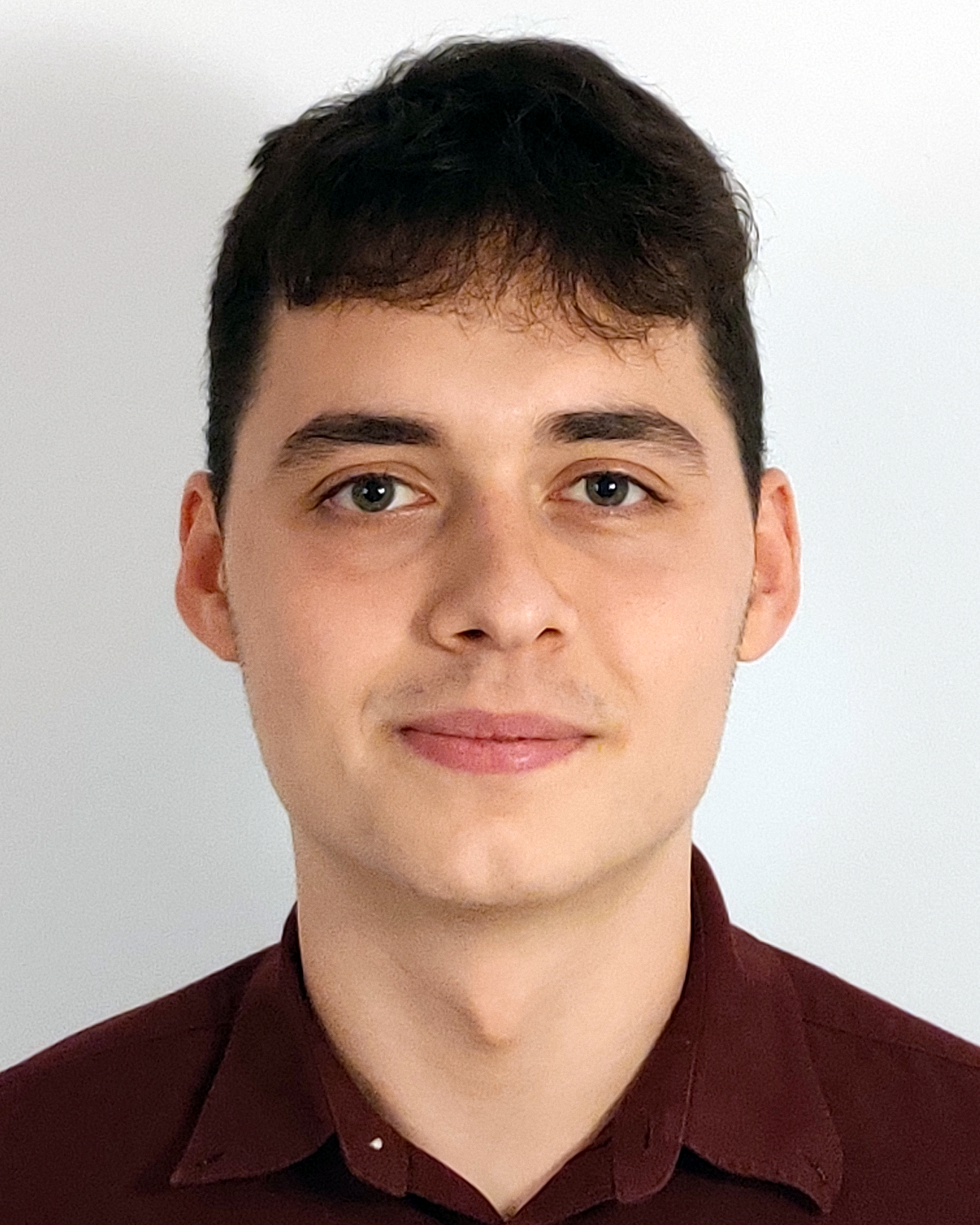}}]{Cezary Adamczyk} is a PhD student at Poznan University of Technology, conducting research in the field of AI/ML utilization in open radio access network radio resource optimization. He works as an OSS Solutions Architect in international telecommunication projects. His prior work related to O\mbox{-}RAN includes a paper proposing a novel Reinforcement Learning algorithm for optimizing radio resources utilization.
\end{IEEEbiography}

\vspace{11pt}

\begin{IEEEbiography}[{\includegraphics[width=1in,height=1.25in,clip,keepaspectratio]{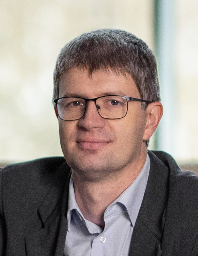}}]{Adrian Kliks} received his postdoctoral degree in technical computer science and telecommunications in February 2019. He works as an associate professor at Poznan University of Technology and has taken part in numerous industrial and commissioned projects. He leads OPUS projects on V2X communication and RISes. An IEEE Senior Member since 2013. In the years 2014-2016 he was the IEEE Membership Development/Web Visibility Chair for the EMEA area. Since 2019 – the editor-in-chief of the Journal of Telecommunications and Information Technology of the Institute of Communications. A vice chair of the IEEE VTS Polish Chapter.
\end{IEEEbiography}

\vfill

\end{document}